\documentclass[12pt]{article}
\newcommand{\be}{\begin{equation}}
\newcommand{\ee}{\end{equation}}
\newcommand{\bea}{\begin{eqnarray}}
\newcommand{\eea}{\end{eqnarray}}

\begin{document}
\begin{titlepage}
\begin{flushright}
Freiburg THEP-99/6\\
gr-qc/9905098
\end{flushright}
\begin{center}

{\large\bf  ORIGIN OF THE INFLATIONARY
           UNIVERSE\protect\footnote{This essay received
 an ``honorable mention'' in the 1999 Essay Competition
 of the Gravity Research Foundation. To appear in Mod. Phys. Lett.~A.}}

\vskip 1cm
{\bf Andrei O. Barvinsky}
\vskip 0.4cm
Theory Department, \\ Levedev Institute and Lebedev Research Center in Physics,\\
Leninsky Prospect 53, Moscow 117924, Russia.\\
E-mail: barvin@td.lpi.ac.ru

\vskip 0.7cm
{\bf Alexander Yu. Kamenshchik}
\vskip 0.4cm
L.D. Landau Institute for Theoretical Physics,\\ Russian Academy of Sciences,\\
Kosygin Street 2, Moscow 117334, Russia.\\
E-mail: kamen@landau.ac.ru
\vskip 0.2cm
Landau Network-Centro Volta,\\ Villa Olmo, Via Cantoni 1,\\
I-22100 Como, Italy.

\vskip 0.7cm
{\bf Claus Kiefer}
\vskip 0.4cm
 Fakult\"at f\"ur Physik, Universit\"at Freiburg,\\
  Hermann-Herder-Stra\ss e 3, D-79104 Freiburg, Germany.\\
 E-mail: Claus.Kiefer@physik.uni-freiburg.de

\end{center}
\vskip1cm
\small
\begin{center}
{\bf Abstract}
\end{center}
\begin{quote}
We give a consistent description of how the inflationary Universe
emerges in quantum cosmology. This involves two steps:
Firstly, it is shown that a sensible probability peak
can be obtained from the cosmological wave function. This is achieved
by going beyond the tree level of the semiclassical expansion.
Secondly, due to decoherence 
interference terms between different
semiclassical branches are negligibly small.
 The results give constraints on the
particle content of a unified theory.
\end{quote}
\normalsize

\end{titlepage}

The standard big-bang model provides a successful scenario
for the evolution of our Universe. Although there is still
some uncertainty about the very early phase, the idea that
the Universe underwent a period of exponential
(``inflationary'') expansion at an early time
(about $10^{-33}\ \mbox{sec}$ after the big bang) is very
promising. Not only does such a scenario avoid some of the
shortcomings of the standard model, it can also give quantitative
predictions for structure formation. In fact, all observed
structure in the Universe can be traced back to quantum fluctuations
during the inflationary era. The predictions are consistent with
the anisotropy spectrum of the cosmic microwave background
radiation observed by the COBE-satellite and earth-based
telescopes.

What is the origin of inflation? Usually, the
{\em no-hair conjecture} is invoked \cite{KT,BG}. This conjecture
states that for a positive (effective) cosmological constant,
a general spacetime approaches locally a De~Sitter metric
for asymptotically late times. The positive cosmological constant
is assumed to arise from particle physics and is here taken 
for granted. The implicit assumption in this conjecture
is that scales originally smaller than the Planck length do not affect
the expansion. This is, however, an assumption about the unknown
quantum theory of gravity. 

This situation has led to the general belief that the origin
of inflation can only be understood within {\em quantum}
cosmology. In the following we shall attempt to give
a precise description of this quantum origin. This might not
be the final answer, but yields a consistent picture that
is based on present knowledge.
We shall, in particular, give answers to the questions:
Can one calculate from quantum cosmology the {\em probability}
of inflation? Moreover, can one quantitatively understand
how the transition from the quantum era to the classical
evolution proceeds?

One might wonder whether these questions can only be answered
after a final quantum theory of gravity -- perhaps superstring theory --
will be available. It is, however, generally assumed that inflation
occurs at a scale about five orders of magnitude below the
Planck scale. This is also indicated by the size $\delta$ of the
anisotropies in the spectrum of the cosmic microwave background,
$\delta\approx10^{-5}$. According to the inflationary scenario,
this should reflect the fact that $m_I/m_P\approx 10^{-5}$, where $m_I$ is
the energy scale of inflation and $m_P=\sqrt{\hbar c/G}$ is the
Planck mass. Since genuine quantum-gravitational effects
are of the order $(m_I/m_P)^2\approx 10^{-10}$
\cite{semi}, an effective
theory of quantum gravity should be sufficient to calculate effects
at this scale. As the classical limit
at such scales should be general relativity, the effective
theory should be canonical quantum gravity to an
excellent approximation.

The central quantity in canonical quantum cosmology is the wave function
$\Psi(a,\varphi,f)$, where $a$ is the scale factor,
$\varphi$ is the field causing inflation (the ``inflaton''),
and $f$ denotes all other degrees of freedom.
The wave function obeys
the Wheeler-DeWitt equation and does not contain any information
about a classical time parameter $t$. Such a time parameter can, however,
be recovered, $\Psi(a,\varphi,f)\rightarrow
\Psi_t(\varphi,f)$,
in a semiclassical approximation with respect to
$m_I/m_P$ \cite{semi}. Since this quantity is small in our case,
the validity of this approximation should be excellent. Then
$a$ and $\varphi$ are not independent, but related through the
semiclassical time $t$; we could alternatively use $a$ as an argument
of $\Psi_t(\varphi,f)$.

The semiclassical approximation can be performed either
as a WKB limit for the Wheeler-DeWitt equation
or as a saddle-point limit on the
path integral. In both ways one finds an approximate classical
spacetime on which quantum degrees of freedom propagate.
To recover an inflationary Universe in the classical limit,
this spacetime is taken to be an approximate {\em De~Sitter} space.
But what boundary conditions should one use? Already Albert Einstein
had to deal with the problems of specifying boundary conditions,
although in ordinary space: ``If it were possible to consider
the world as a continuum that is closed in all spatial directions,
no such boundary conditions would be necessary at all''.\footnote{``Wenn
es n\"amlich m\"oglich w\"are, die Welt als ein nach seinen
r\"aumlichen Erstreckungen geschlossenes Kontinuum anzusehen,
dann h\"atte man \"uberhaupt keine derartigen
Grenzbedingungen n\"otig.'' \cite{Ein}}
In quantum cosmology, a similar idea is contained in the
so-called no-boundary proposal \cite{HH}: One should sum in the
path integral only over such manifolds that have {\em no} initial
boundary. In the semiclassical approximation, this gives rise
to the {\em De~Sitter instanton}: Classical De~Sitter space is
attached to half of a euclidean four-sphere with radius
\be H^{-1}(\varphi)\equiv \frac{3m_P^2}{8\pi V(\varphi)}\ , \ee
where $V(\varphi)$ is the inflationary potential, and
$H$ is the Hubble parameter.

There also exist other sensible boundary conditions including
the so-called tunneling one \cite{boun}. Although they do not have such a
geometrical interpretation, in the semiclassical approximation
they just correspond to the choice of a different
WKB solution.

To discuss both the probability for inflation and the emergence
of classical properties, the reduced density matrix for
scale factor and inflaton should be investigated. This density matrix
is calculated from the full quantum state upon integrating out the
degrees of freedom $f$,
\be \rho_t(\varphi,\varphi')=\int{\cal D}f\
       \Psi_t^*(\varphi',f)\Psi_t(\varphi,f)\ . \ee
To calculate the probability one has to set $\varphi'=\varphi$.
In earlier work, the saddle-point approximation was only performed
up to the highest, tree-level, approximation \cite{HH,boun}.
This yields
\be \rho(\varphi,\varphi)=\exp[\pm I(\varphi)]\ , \ee
where $I(\varphi)=-3m_P^4/8V(\varphi)$. The lower sign corresponds
to the no-boundary condition, while the upper sign corresponds to
the
tunneling condition \cite{boun}. The problem with (3) is that $\rho$ is not
normalisable: Scales $m_I\equiv H(\varphi)>m_P$ contribute significantly
and thus spoil the original idea of the no-hair conjecture. Results
based on tree-level approximations can thus not be trusted.

The situation is considerably improved if
loop effects are taken into account \cite{norm,qsi,BKM}.
They are incorporated by the loop effective action $\Gamma_{loop}$
which is calculated on the
De-Sitter instanton. In the limit of large $\varphi$
(that is relevant for investigating normalisability)
this yields in the one-loop approximation
\be \Gamma_{loop}(\varphi)\vert_{H\to\infty}
        \approx Z\ln\frac{H}{\mu}\ , \ee
where $\mu$ is a renormalisation mass parameter, and $Z$ is the
anomalous scaling.
Instead of (3) one has now \cite{norm}
\bea \rho(\varphi,\varphi)&\approx& \ H^{-2}(\varphi)
     \exp\left(\pm I(\varphi)-\Gamma_{loop}(\varphi)\right)
     \nonumber\\
    &\approx& \ \exp\left(\pm\frac{3m_P^4}{8V(\varphi)}\right)
    \varphi^{-Z-2} \ . \eea
This density matrix is normalisable provided $Z>-1$.
This in turn leads to reasonable constraints on the
particle content of the theory \cite{norm,qsi}.

One can easily obtain a probability peak for the initial
value of the inflaton, $\varphi_I$, at the onset of inflation
in a tunneling model with nonminimal coupling \cite{qsi,BKM}:
$\varphi_I\approx 0.03m_P$, with a dispersion
$\Delta\varphi\approx 10^{-7}m_P$, and a corresponding
Hubble parameter $H(\varphi_I)\approx 10^{-5}m_P$.
The relative width
\be \frac{\Delta\varphi}{\varphi_I}\approx
     \frac{\Delta H}{H}\approx 10^{-5} \ee
corresponds to the observed anisotropies in the cosmic microwave
background. This approach also works for open inflation with 
no-boundary initial condition \cite{Bar}. We emphasise that no
anthropic principle is needed to obtain these results, in contrast
to earlier tree-level calculations.

Since quantum theory by itself does not yet yield
classical ensembles, it would be premature to interpret
the above results as giving by itself the probabilities
for inflationary universes with different Hubble parameters.
This can only be done after a quantative understanding of the
quantum-to-classical transition has been gained.
How can this be achieved?

It is now quite generally accepted that classical properties for
a subsystem emerge from the irreversible interaction of this
quantum system with its natural environment -- a process called
{\em decoherence} \cite{deco}. But what constitutes system and
environment in quantum cosmology where there is no external
measuring agency? It was suggested in \cite{Zeh, Kie}
to consider the global degrees of freedom $a$ and $\varphi$
as the ``relevant'' variables which are decohered by ``irrelevant''
degrees of freedom such as density fluctuations, gravitational
waves, or other fields. The latter comprise the variables $f$ that
are integrated over in (2).

Since information about interference terms is contained
in the non-diagonal elements of the reduced density matrix,
one has to evaluate (2) for $\varphi\neq\varphi'$
(or alternatively, $a\neq a'$, which we use
for illustration below). The resulting expressions are,
however, ultraviolet-divergent and must be regularised. They
also depend on the parametrisation of quantum fields.
This was investigated in detail for bosons in \cite{BKKM}
and for fermions in \cite{BKK}. Unfortunately, standard regularisation
schemes such as dimensional regularisation do not work since they spoil
a crucial property of the density matrix -- its boundedness.
We have thus put forward the physical
principle that decoherence should be absent in the absence of
particle creation since decoherence is an irreversible process.
There should, in particular, be no decoherence
for static spacetimes. This has guided us to use a certain
conformal reparametrisation for bosonic fields and a
certain Bogoliubov transformation for fermionic fields, which
renders the decoherence effects finite.

We have calculated for the above semiclassical
De~Sitter background the decoherence factor $D_t(\varphi,\varphi')$
that multiplies the expression for the reduced density matrix in the
absence of interactions. Because there is no particle creation
if the ``environmental'' fields are massless and conformally
invariant, $D_t(\varphi,\varphi')=1$ and decoherence is absent. This
is no longer true for other fields. Taking a massive scalar field
with mass $m$, for example,
one finds for the absolute value of the decoherence factor
in the $a$-representation
\be |D_t(a,a')|\approx \exp\left(-\frac{\pi m^3a}{128}
        (a-a')^2\right) \ . \ee
Interference terms between different $a=\cosh Ht/H$
and $a'=\cosh H't/H'$, $H=H(\varphi),\;H'=H(\varphi')$, are
therefore suppressed, and the effect becomes more efficient
with increasing $t$. The result for gravitons is similar,
with the mass $m$ being replaced by the Hubble parameter $H$.
For massive fermions, the power of the mass is
$m^2$ instead of $m^3$.

It becomes clear from these examples that the Universe
acquires classical properties after the onset of the
inflationary phase, and it
makes sense to speak of a probability distribution
for classical universes. ``Before'' this phase, the Universe
was in a timeless quantum state which does not possess any
classical properties. Viewed backwards, different semiclassical
branches would meet and interfere to form this timeless
quantum state. After the background has become classical,
the stage is set for other degrees of freedom to
assume classical properties, such as for the primordial
fluctuations that arise in the inflationary Universe
and lead to the observed anisotropies of the
microwave background radiation
\cite{KPS}.

\section*{Acknowledgements}
\hspace{\parindent}
The work of A.B. was partially supported by RFBR under the grant No
99-02-16122. The work of A.K. was partially supported by RFBR under the
grant 99-02-18409 and under the grant for support of leading
scientific schools 96-15-96458. A.B. and A.K. kindly acknowledge
financial support by the DFG grants 436 RUS 113/333/4 during their
visit to the University of Freiburg in autumn~1998.

\end{document}